# Cosmological model-independent constraints on the baryon fraction in the IGM from fast radio bursts and supernovae data

Thais Lemos[1,a], Rodrigo S. Gonçalves[1,2,b], Joel C. Carvalho[1,c], Jailson S. Alcaniz[1,d]

[1] Observatório Nacional, Rio de Janeiro, RJ 20921-400, Brazil
[2] Departamento de Física, Universidade Federal Rural do Rio de Janeiro, Seropédica, Rio de Janeiro 23897-000, Brazil

**Abstract** Fast Radio Bursts (FRBs) are millisecond-duration radio transients with an observed dispersion measure ($DM$) greater than the expected Milky Way contribution, which suggests that such events are of extragalactic origin. Although some models have been proposed to explain the physics of the pulse, the mechanism behind the FRBs emission is still unknown. From FRBs data with known host galaxies, the redshift is directly measured and can be combined with estimates of the $DM$ to constrain the cosmological parameters, such as the baryon number density and the Hubble constant. However, the poor knowledge of the fraction of baryonic mass in the intergalactic medium ($f_{IGM}$) and its degeneracy with the cosmological parameters impose limits on the cosmological application of FRBs. In this work we present a cosmological model-independent method to determine the evolution of $f_{IGM}$ combining the latest FRBs observations with localized host galaxy and current supernovae data. We consider constant and time-dependent $f_{IGM}$ parameterizations and show, through a Bayesian model selection analysis, that a conclusive answer about the time-evolution of $f_{IGM}$ depend strongly on the $DM$ fluctuations due to the spatial variation in cosmic electron density ($\delta$). In particular, our analysis show that the evidence varies from strong (in favor of a growing evolution of $f_{IGM}$ with redshift) to inconclusive, as larger values of $\delta$ are considered.

## 1 Introduction

Fast Radio Bursts (FRBs) are energetic radio transients with duration time of the order of millisecond and typical radiation frequency of ∼ GHz [1–4]. Although some models have been proposed to explain the physics of the pulse and some of these events are associated with magnetars [5], the origin of the FRBs emission remains unknown [6]. Since the value of the dispersion measure observed (DM) is greater than the one expected from the Milky Way contribution, FRBs are thought to be extragalactic events or even of cosmological origin [7]. The origin of the pulse is confirmed when it is possible to identify its host galaxy and, consequently, its redshift.

The first FRB was discovered by the Parkes Telescope in 2007 and was named FRB 010724 [1]. After that, more than one hundred FRBs have been discovered [4,8]. These events can be divided in two groups according to if they are repeating or nonrepeating. Apparently, most of the bursts found are nonrepeating [9,10]. If one can identify the host galaxy of the bursts, we can use the dispersion measure ($DM$) versus redshift ($DM-z$) relation of these events as a tool to study the underlying cosmology. In fact, FRBs have been used to constrain cosmological parameters [11,12], such as the Hubble parameter $H(z)$ [13] and Hubble constant $H_0$ [14,15], to probe the anisotropic distribution of baryon matter in Universe [16], as well as to constrain the fraction of baryon mass in the intergalactic medium (IGM) [17–19].

One issue that restricts the application of FRBs in cosmology is the uncertainties on the evolution of the fraction of baryon mass in the IGM ($f_{IGM}$) and its degeneracy with the cosmological parameters. In this concern, some studies have been performed to discuss the baryon distribution in the IGM using both numerical simulations [20–22] and observations [23–25]. For instance, in Ref. [22] the authors performed numerical simulations and found that about 90% of the baryons produced by the Big Bang are contained within the IGM at $z \geq 1.5$ (i.e., $f_{IGM} \approx 0.9$) whereas in Ref. [24], the baryons existent in the collapsed phase at $z \geq 0.4$ represent $18 \pm 4$% or, equivalently, $f_{IGM} \approx 0.82$. From these results one may naively infer that the $f_{IGM}$ grows with red-shift. Other recent analyses have pointed out that if a sample of FRBs can be localized, so their luminosity distance $d_L$ can be determined [12,17,26].

<sup>a</sup> e-mail: thaislemos@on.br (corresponding author)
<sup>b</sup> e-mail: rsg_goncalves@ufrrj.br
<sup>c</sup> e-mail: jcarvalho@on.br
<sup>d</sup> e-mail: alcaniz@on.br

In this paper, we propose a new cosmological model-independent method to constrain a possible evolution of $f_{IGM}(z)$ directly from observations of FRBs dispersion measure $DM(z)$ and $d_L$ from type Ia supernovae (SNe) data. In our analysis, we use a subsample of 17 FRBs with known redshifts [27–37] along with the Pantheon SNe catalogue [38]. We consider both constant and time-dependent parameterizations for $f_{IGM}$ and discuss the observational viability of them through a Bayesian model selection analysis. We organized this paper as follows. In Sect. 2 we introduce our method to study the evolution of $f_{IGM}$ with redshift. The data sets used in the analysis and their application are presented and discussed in Sect. 3. We present our main results in Sect. 4. The role of the DM fluctuations in the determination of the $f_{IGM}$ evolution is discussed in Sect. 5. We end the paper in Sect. 6 by presenting our main conclusions.

## 2 A new method to determine the baryon fraction

### 2.1 Dispersion measure

The observed $DM$ of a FRB is a combination of several components [26,39]:

$$DM_{obs}(z) = DM_{MW} + DM_{IGM}(z) + DM_{host}(z), \quad (1)$$

where the subscripts MW, IGM and host denote contributions from the Milky Way, IGM, and the FRB host galaxy, respectively. The observed $DM$ of a FRB is directly measured from the corresponding event while the $DM$ of the Milky Way has a contribution from the Milky Way interstellar medium (ISM) and from the Milky Way halo, estimated by the relation $DM_{MW} = DM_{MW,ISM} + DM_{MW,halo}$ [40]. $DM_{MW,ISM}$ can be well constrained using models of the ISM galactic electron distribution in the Milky Way from pulsar observations [41–43] whereas the Milky Way halo contribution is not well constrained yet. In our analysis, we follow [40] and assume $DM_{MW,halo} = 50$ pc/cm$^3$.

Subtracting the Galaxy contribution from the observation of $DM$ we define the observed extragalactic $DM$ as

$$DM_{ext}(z) \equiv DM_{obs}(z) - DM_{MW}, \quad (2)$$

so that, using Eq. (1), the theoretical extragalactic $DM$ can be calculated as

$$DM_{ext}^{th}(z) \equiv DM_{IGM}(z) + DM_{host}(z), \quad (3)$$

where both terms on the right hand side are described as follows.

The redshift evolution of $DM_{host}(z)$ is given by [39,44]:

$$DM_{host}(z) = \frac{DM_{host,0}}{(1+z)}, \quad (4)$$

where the $(1 + z)$ factor accounts for the cosmic dilation. The host galaxy contribution, $DM_{host,0}$, is a poorly known parameter, as it depends on the type of the galaxy, the relative orientations of the FRBs source with respect to the host and source, and the near-source plasma [45]. Therefore, the host galaxy contribution $DM_{host,0}$ will be considered a free parameter in our analysis. On the other hand, the average dispersion measure from IGM can be written as function of the redshift as [39]

$$DM_{IGM}(z) = \frac{3c\Omega_b H_0^2}{8\pi G m_p} \int_0^z \frac{(1+z')f_{IGM}(z')\chi(z')}{H(z')}dz', \quad (5)$$

where $c$ is the speed of light, $\Omega_b$ is the present-day baryon density parameter, $H_0$ is the Hubble constant, $G$ is the gravitational constant, $m_p$ is the proton mass, $f_{IGM}(z)$ is the baryon fraction in the IGM, $H(z)$ is the Hubble parameter at redshift $z$ and the free electron number fraction per baryon is given by

$$\chi(z) = Y_H \chi_{e,H}(z) + Y_{He}\chi_{e,He}(z). \quad (6)$$

The terms $Y_H = 3/4$ and $Y_{He} = 1/4$ are the mass fractions of hydrogen and helium, respectively, while $\chi_{e,H}(z)$ and $\chi_{e,He}(z)$ are the ionization fractions of hydrogen and helium, respectively. At $z < 3$ hydrogen and helium are fully ionized ($\chi_{e,H}(z) = \chi_{e,He}(z) = 1$) [22,46], so that we have $\chi(z) = 7/8$. From the above equations, one can constrain a possible evolution of the baryon fraction by modelling both $DM_{host,0}$ and $DM_{IGM}$ and comparing the theoretical predictions with the observed values of $DM_{ext}$.

### 2.2 $f_{IGM}(z)$ from FRB and SNe observations

As mentioned earlier, one of the aspects that restricts the application of FRBs in cosmology is the uncertainties on the evolution $f_{IGM}$ with redshift. In order to investigate this matter further, we assume in our analysis two parameterizations for this quantity:

$$f_{IGM} = f_{IGM,0}, \quad (7a)$$

$$f_{IGM} = f_{IGM,0} + \alpha \frac{z}{1+z}. \quad (7b)$$

The parameter $f_{IGM,0}$ is the present value of $f_{IGM}$ whereas $\alpha$ quantifies a possible evolution of $f_{IGM}$. In our analysis both are free parameters and since $f_{IGM}$ is understood to be an increasing function of the redshift, $\alpha$ assumes only positive values ($\alpha \geq 0$). Hereafter, we explicit the $DM_{IGM}$ expression for both cases.

Considering the general case in which $f_{IGM}(z)$ is a function of redshift, one can calculate Eq. (5) by parts:

$$DM_{IGM}(z) = Af_{IGM}(z)\frac{d_L(z)}{c} - \int_0^z \frac{d_L(z')}{(1+z')c} \\ \times Af_{IGM}(z')dz' - A\int_0^z \frac{d_L(z')}{c} f'_{IGM}(z')dz', \quad (8)$$

where $A = \frac{21c\Omega_b H_0^2}{64\pi G m_p}$, $f'_{IGM}(z) = \frac{df_{IGM}(z)}{dz}$ and

$$d_L(z) = c(1+z)\int_0^z \frac{dz'}{H(z')} \quad (9)$$

is the luminosity distance. Now replacing parameterization (7b) in the above expression we obtain

$$DM_{IGM}(z) = A\left(f_{IGM,0} + \alpha\frac{z}{1+z}\right)\frac{d_L(z)}{c} \\ - A\left(f_{IGM,0} + \alpha\right)\int_0^z \frac{d_L(z')}{c(1+z')}dz'. \quad (10)$$

For the constant case (7a), we follow the same steps above and find

$$DM_{IGM}(z) = Af_{IGM,0}\left[\frac{d_L(z)}{c} - \int_0^z \frac{d_L(z')}{(1+z')c}dz'\right]. \quad (11)$$

Note that the last term of Eqs. (10) and (11) are equal and can be numerically solved as (see [47]):

$$\int_0^z \frac{d_L(z')}{(1+z')c}dz' = \frac{1}{2c}\sum_{i=1}^N (z_{i+1} - z_i) \\ \times \left[\frac{d_L(z_{i+1})}{(1+z_{i+1})} + \frac{d_L(z_i)}{(1+z_i)}\right]. \quad (12)$$

Therefore, using estimates of $d_L(z)$ from SNe observations, it is possible to constrain the evolution of $f_{IGM}(z)$ with redshift from the above expressions.

## 3 Data and methodology

In order to discuss a possible evolution of the baryon fraction, we use observational data for the dispersion measures and luminosity distance. The former is obtained directly from FRBs measurements whereas the latter comes from SNe observations.

Currently, there are 19 FRBs events with localised host galaxy and redshifts (for details of FRBs catalogue[1] see [4] and for host database[2] see [30]). In our analysis, we use a sample of 16 FRBs within the redshift interval $0.0337 \leq z \leq 0.66$, which constitutes the most up-to-date FRB data set currently available [27–35,37]. Our subsample excludes the repeating burst FRB 20200120E [48] at $z = -0.0001$, observed in the direction of M81, the FRB 20181030A [49] since there is no SNe in the Pantheon catalogue near its redshift ($z = 0.0039$), and the FRB 190614D whose redshift estimate lies in the interval $0.4 \lesssim z \lesssim 0.75$ (68% confidence interval), and can be associated with two host galaxies [36]. The main properties of these 16 FRB events are shown in Table 1, namely: redshift, $DM_{MW,ISM}$, $DM_{obs}$ and observed $DM$ error of all localised FRBs. The values of $DM_{MW,ISM}$ are estimated from the NE2001 model [42].

From Table 1, we can calculate our observational quantity, $DM_{ext}$, using Eq. (2), whose uncertainty is given by

$$\sigma_{ext}^2 = \sigma_{obs}^2 + \sigma_{MW}^2, \quad (13)$$

where the average galactic uncertainty $\sigma_{MW}$ is assumed to be 10 pc/cm$^3$ [50].

In order to obtain measurements of $d_L(z)$, we use the distance moduli ($\mu(z)$) data obtained from current SNe Ia observations. This quantity is related to $d_L(z)$ by

$$\mu(z) = m_B - M_B = 5\log_{10}\left[\frac{d_L(z)}{1\text{Mpc}}\right] + 25, \quad (14)$$

where $m_B$ is the apparent magnitude of SNe, and, in our analysis, we fix the absolute peak magnitude at $M_B = -19.214 \pm 0.037$ mag, as given by [51]. The data set used for SNe is the Pantheon catalogue [38], which comprises 1048 SNe within the redshift range $0.01 < z < 2.3$. In order to work with the equations derived in the previous section, we perform a Gaussian Process (GP) reconstruction of the Pantheon data to obtain estimates of $d_L(z)$ at the same redshifts of the FRBs (for details of GP reconstructions we refer the reader to [52,53] and references therein).[3]

Summarizing, the steps of our analysis are the following: first, we calculate $DM_{ext}(z_i)$ observed and $\sigma_{ext}(z_i)$ using the FRBs dataset. Second, the luminosity distance is calculated at the same $DM_{ext}$ redshift, using the GP reconstruction of Pantheon catalogue. The integral given by Eq. (12) is then calculated with the SNe data, considering that the redshift limit of the sum ($z_L$) must be equal to the redshift of the FRB ($z_L = z_i$). Finally, we use the Monte Carlo Markov Chain (MCMC) method to fit the free parameters of our analysis, i.e., $f_{IGM}$ and $DM_{host,0}$ in the constant case of Eq. (7a) and $f_{IGM,0}$, $\alpha$ and $DM_{host,0}$ for the time-dependent parameterization (7b). The MCMC analysis is performed with the emcee sample [54], and to be consistent with our choice of $M_B$ in Eq. (14) – since we are also interested in model-independent approach – we adopt the value of the Hubble constant from the SH0ES collaboration, $H_0 = 74.03 \pm 1.4$ km s$^{-1}$ Mpc$^{-1}$ [51]. We also assume $\Omega_b h^2 = 0.02235 \pm 0.00037$, as reported by [55].

---

[1] http://www.frbcat.org.

[2] https://frbhosts.org/.

[3] An alternative approach is to define a redshift interval centered at the redshifts of each FRB and calculate the average values of $d_L(z)$ from the SNe data within the interval. We verified this approach and obtained results (not shown here) very similar to the ones derived through the GP reconstruction.

**Table 1** Properties of FRB with known host galaxies

| Name | Redshift $z$ | $DM_{MW,ISM}$ [pc/cm$^3$] | $DM_{obs}$ [pc/cm$^3$] | $\sigma_{obs}$ [pc/cm$^3$] | References |
| --- | --- | --- | --- | --- | --- |
| FRB 180916B | 0.0337 | 200.0 | 348.8 | 0.2 | [27] |
| FRB 201124A | 0.098 | 123.2 | 413.52 | 0.5 | [28] |
| FRB 190608B | 0.1178 | 37.2 | 338.7 | 0.5 | [29] |
| FRB 200430A | 0.16 | 27.0 | 380.25 | 0.4 | [30] |
| FRB 121102A | 0.19273 | 188.0 | 557.0 | 2.0 | [31] |
| FRB 191001A | 0.234 | 44.7 | 506.92 | 0.04 | [30] |
| FRB 190714A | 0.2365 | 38.0 | 504.13 | 2.0 | [30] |
| FRB 20191228A | 0.2432 | 33.0 | 297.5 | 0.05 | [32] |
| FRB 190102C | 0.291 | 57.3 | 363.6 | 0.3 | [33] |
| FRB 180924B | 0.3214 | 40.5 | 361.42 | 0.06 | [34] |
| FRB 20180301A | 0.3305 | 152.0 | 536.0 | 8.0 | [32] |
| FRB 20200906A | 0.3688 | 36.0 | 577.8 | 0.02 | [32] |
| FRB 190611B | 0.378 | 57.83 | 321.4 | 0.2 | [30] |
| FRB 181112A | 0.4755 | 102.0 | 589.27 | 0.03 | [35] |
| FRB 190711A | 0.522 | 56.4 | 593.1 | 0.4 | [30] |
| FRB 190523A | 0.66 | 37.0 | 760.8 | 0.6 | [30,37] |

## 4 Results

In Fig. 1, we show the posterior probability density function and $1 - 2\sigma$ constraint contours of the free parameters ($f_{IGM,0}$, $\alpha$, $DM_{host,0}$) for the constant case (left Panel) and the time-dependent parameterization (right Panel). We also present in Table 2 the results for the baryon fraction for both cases. For the constant case, we obtain $f_{IGM,0} = 0.764 \pm 0.013$ and the estimate for the host galaxy contribution $DM_{host,0} = 158.1 \pm 5.4$ pc/cm$^3$, both at 1$\sigma$ level. The result for the baryon fraction is in good agreement with previous results obtained from observations [23–25,56] and numerical simulations [14,20,21]. For the time-dependent case, we obtain $f_{IGM,0} = 0.483 \pm 0.066$ (1$\sigma$) for the present value of the baryon fraction and $\alpha = 1.21 \pm 0.28$ (1$\sigma$). We also estimate the host galaxy contribution at $DM_{host,0} = 190.1 \pm 9.1$ pc/cm$^3$ (1$\sigma$). We note that the values of $f_{IGM,0}$ and $\alpha$ do not show agreement with other recent studies that used the same parameterization (7b) – see e.g. [18,57]. We believe that such discrepancy may be primarily related to the fact that these works do not consider the contribution of the host galaxy $DM_{host}$ as a free parameter in their analyses, as well as to the more up-to-date FRB data used in our analysis.

Another important aspect of the above results concerns the observational evidence for a evolution of the baryon fraction $f_{IGM}$ with redshift. In order to evaluate the two cases studied and quantify such evidence, we perform a Bayesian model comparison. This kind of analysis offers a way to assess if the extra complexity of a given model or parameterization (here represented by the parameter $\alpha$) is required by the data, preferring the model that describes the data well over a large fraction of their prior volume (see e.g. [58,59] for a detailed discussion).

By defining the evidence as the marginal likelihood of the models, we calculate the Bayes' factor $B_{ij}$:

$$B_{ij} = \frac{\mathcal{E}_i}{\mathcal{E}_j}, \quad (15)$$

where $\mathcal{E}_i$ and $\mathcal{E}_j$ correspond to the evidence of parameterizations $\mathcal{P}_i$ and $\mathcal{P}_j$, respectively. We adopted the Jeffreys' scale [60] to interpret the values of $\ln B_{ij}$ for the reference parameterization $\mathcal{P}_j$: $\ln B_{ij} = 0 - 1$, $\ln B_{ij} = 1 - 2.5$, $\ln B_{ij} = 2.5 - 5$, and $\ln B_{ij} > 5$ indicate, respectively, an inconclusive, weak, moderate and strong preference of the parameterization $\mathcal{P}_i$ with respect to $\mathcal{P}_j$. Negative values of $\ln B_{ij}$ mean preference in favour of $\mathcal{P}_j$.

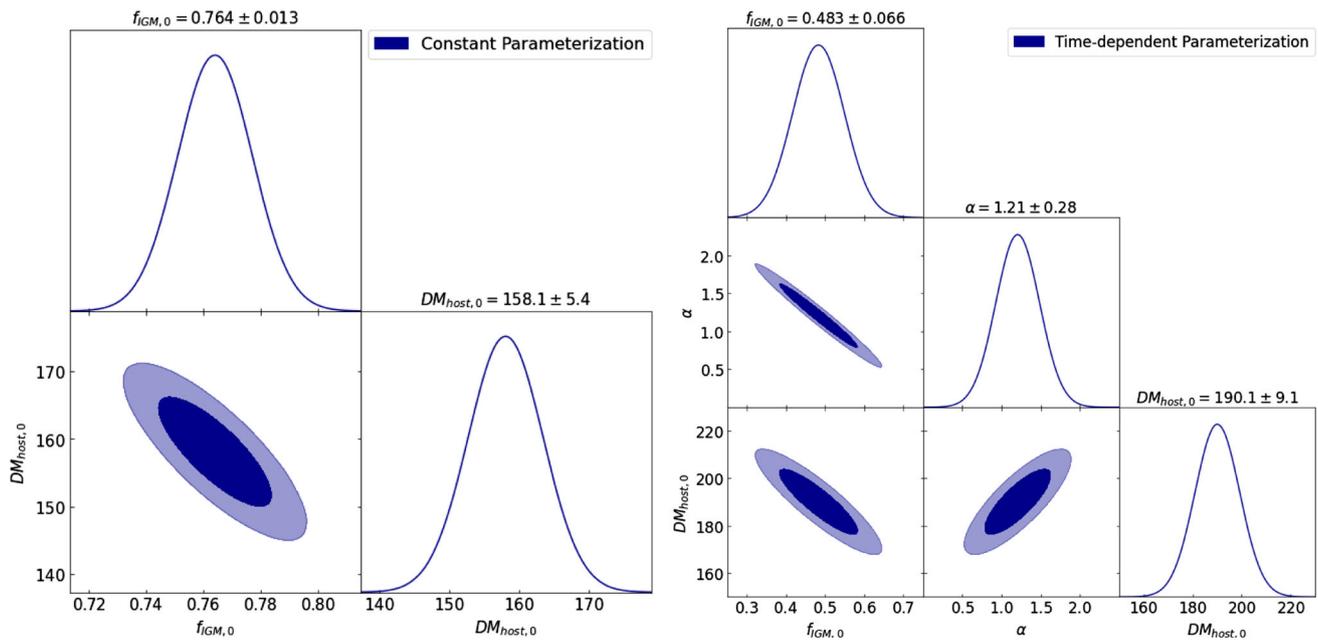

**Fig. 1** Left: Constraints on the baryon fraction $f_{IGM}$ and the mean host galaxy contribution of dispersion measure $DM_{host,0}$ considering the constant case (7a). Right: Constraints on the present-day baryon fraction $f_{IGM,0}$, $\alpha$ and the mean host galaxy contribution of dispersion measure $DM_{host,0}$ for the time-dependent parameterization (7b)

**Table 2** Estimates of the parameters $f_{IGM,0}$, $\alpha$ and $DM_{host,0}$ for the two parameterizations considered in the analysis

|  | $f_{IGM,0}$ | $\alpha$ | $DM_{host,0}$ [pc/cm$^3$] |
| --- | --- | --- | --- |
| Constant | $0.764 \pm 0.013$ | – | $158.1 \pm 5.4$ |
| Time-dependent | $0.483 \pm 0.066$ | $1.21 \pm 0.28$ | $190.1 \pm 9.1$ |

We use the MultiNest algorithm [61–63] to compute the Bayesian evidence ($\ln \mathcal{E}$) and then calculate the Bayes' factor. Adopting the constant case (7a) as reference, we obtain $\ln \mathcal{E}_j = -565.349$ and $\ln \mathcal{E}_i = -557.032$ for the reference and time-dependent cases, respectively, which results in $\ln B_{ij} = 8.32$. Such a result indicates a strong evidence in favor of the time-dependent parameterization (7b) with respect to the constant case (7a), with the interval of values of the parameters $f_{IGM,0}$, $\alpha$ and $DM_{host,0}$ given by Table 2. For completeness, we also show in Fig. 2 the 3$\sigma$ envelope for the evolution of $DM_{ext}$ with redshift (Eq. 3) considering both parameterizations. The analysis above clearly shows the potential of the method proposed here to probe a possible evolution $f_{IGM}$ with redshift.

## 5 DM fluctuations

In the previous sections we presented and applied our method to constrain the $f_{IGM}$ evolution without considering the $DM$

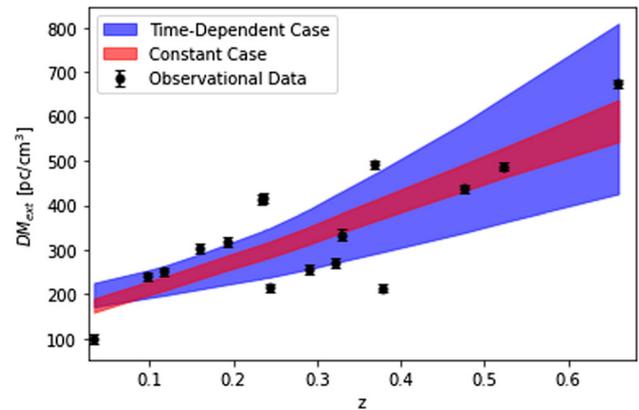

**Fig. 2** The 3$\sigma$ envelope for the evolution of $DM_{ext}$ with redshift considering the constant (red) and time-dependent (blue) parameterizations

fluctuations ($\delta$) due to the spatial variation in cosmic electron density (see e.g. [65] for a detailed discussion on the $DM$ fluctuations). Such fluctuations are not currently well determined by observations and can be treated as a probability distribution or as fixed value in the statistical analyses [15,40,64]. In order to assess the impact of theses fluctuations in the results presented in Sect. 4, we redo the analysis of Sect. 3 considering three different values for this quantity, $\delta = 10, 50, 100$ pc/cm$^3$, being the latter in agreement with the results reported in [65].

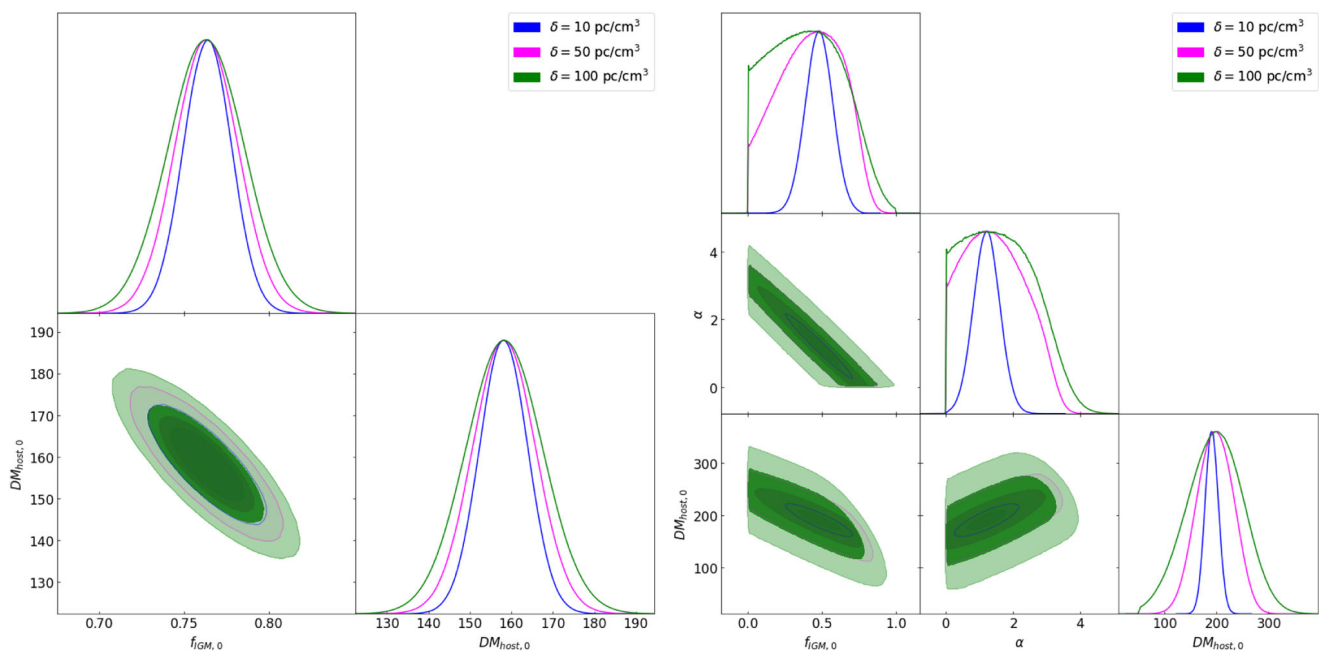

**Fig. 3** The same as in Fig. 1 considering fluctuations in the FRB's $DM$

The results of our analysis are displayed in Fig. 3 and Table 3. Figure 3 shows the posterior probability density function and 1–2$\sigma$ contours on the parametric spaces for different values of the $DM$ fluctuations. The quantitative results of the analysis, displayed in Table 3, show the impact of the DM fluctuations in the determination of the $f_{IGM}$ evolution, as the evidence varies from moderate (in favor of a growing evolution of $f_{IGM}$ with redshift) to inconclusive and inconclusive for $\delta = 10, 50, 100$ pc/cm$^3$, respectively. Therefore, differently from the results presented in Sect. 3 ($\delta = 0$), these results show that a conclusive answer about the time-evolution of $f_{IGM}$ depend strongly on the $DM$ fluctuations and cannot be achieved from the current FRB observational data.

## 6 Conclusions

A proper understanding of the evolution of the baryon fraction in the intergalactic medium ($f_{IGM}$) is one of the main issues concerning the use of FRB observations as a cosmological test. In this paper, following previous studies (see e.g. [17]) but proposing a completely different approach, we presented a cosmological model-independent method for estimating the evolution of $f_{IGM}$ and the local value of the host-galaxy DM using current measurements of luminosity distance from SNe observations and dispersion measures for FRBs.

Following the semi-analytical method described in Sect. 2, in which $DM_{IGM}(z)$ is written in terms of $d_L(z)$, we investigated the current constraints on the baryon fraction considering two different behaviours for this quantity, which are expressed by the constant and time-dependent parameterizations given by Eqs. (7). We used 16 FRB observations, the most up-to-date data currently available, and a GP reconstruction of 1048 SNe from the Pantheon catalogue to perform a MCMC analysis considering the host galaxy contribution for the dispersion measure $DM_{host,0}$ as a free parameter and no $DM$ fluctuation ($\delta = 0$). For the constant case, we found $f_{IGM,0} = 0.764 \pm 0.013$ and $DM_{host,0} = 158.1 \pm 5.4$ pc/cm$^3$ (1$\sigma$) whereas for the time-dependent case we obtained $f_{IGM,0} = 0.483 \pm 0.066$, $\alpha = 1.21 \pm 0.28$, and $DM_{host,0} = 190.1 \pm 9.1$ pc/cm$^3$ at 1$\sigma$ level. In order to evaluate the observational viability of the two cases considered in the analysis we also performed a Bayesian model comparison. Such results showed a strong evidence (ln $B_{ij} = 8.32 \pm 0.01$ at 1$\sigma$) in favor of a increasing evolution of $f_{IGM}$ with redshift.

The results are much less conclusive when the $DM$ fluctuations due to the spatial variation in cosmic electron density are considered in the analysis. In this case, we considered three values of $\delta$ and showed that the strong evidence in favor of a growing evolution of $f_{IGM}$ with redshift obtained in Sect. 3 ($\delta = 0$ pc/cm$^3$) changes to moderate ($\delta = 10$ pc/cm$^3$) and inconclusive ($\delta = 50$ and $100$ pc/cm$^3$). These results clearly show the impact of $DM$ fluctuations in the determination of the $f_{IGM}$ evolution. They also reinforce the interest in searching for a larger sample of FRBs and the need for a better understanding of their physical properties.

**Table 3** Estimates of the $f_{IGM}$ parameters for different values of the $DM$ fluctuations

| $\delta$ [pc/cm$^3$] | $f_{IGM,0}$ | $\alpha$ | $DM_{host,0}$ [pc/cm$^3$] | $\ln \mathcal{E}_i$ | $\ln B_{ij}$ |
| --- | --- | --- | --- | --- | --- |
| 10  | $0.76 \pm 0.02$ | – | $158.3 \pm 7.5$ | $-286.458 \pm 0.007$ | – |
| 50  | $0.76 \pm 0.07$ | – | $158.3 \pm 30.0$ | $-25.328 \pm 0.005$ | – |
| 100 | $0.76 \pm 0.13$ | – | $162.0 \pm 50.0$ | $-7.917 \pm 0.003$ | – |
| 10  | $0.48 \pm 0.09$ | $1.21 \pm 0.39$ | $190.44 \pm 12.70$ | $-282.452 \pm 0.007$ | $4.006 \pm 0.007$ |
| 50  | $0.43 \pm 0.22$ | $1.22 \pm 0.39$ | $197.22 \pm 35.26$ | $-24.757 \pm 0.004$ | $0.571 \pm 0.006$ |
| 100 | $0.40 \pm 0.26$ | $1.56 \pm 1.08$ | $196.56 \pm 53.62$ | $-7.670 \pm 0.004$ | $0.247 \pm 0.005$ |

Finally, it is worth mentioning that current and planned observational programs, such as the Canadian Hydrogen Intensity Mapping Experiment (CHIME) [8] are expected to detect several thousands of FRBs in the next years. These observations will improve significantly the constraints on $f_{IGM}$ from the method proposed here, bringing important information about the matter distribution in the universe as well as demonstrating the potential of FRBs observations for precision measurements of cosmological parameters.

**Acknowledgements** TL thanks the financial support from the Coordenação de Aperfeiçoamento de Pessoal de Nível Superior (CAPES). JSA is supported by Conselho Nacional de Desenvolvimento Científico e Tecnológico (CNPq 310790/2014-0) and Fundação de Amparo à Pesquisa do Estado do Rio de Janeiro (FAPERJ) Grant 259610 (2021).

**Data Availability Statement** This manuscript has associated data in a data repository. [Authors' comment: The data used in this paper can be found in refs. [27–35,37,38].]